\documentclass[twocolumn,prl]{revtex4}

 \usepackage[dvips]{graphicx}

\usepackage{amssymb}

\begin{document}

\title{Cooper-pair qubit and Cooper-pair electrometer in one device}


\author{A.~B.~Zorin}

\address{Physikalisch-Technische Bundesanstalt, Bundesallee 100, D-38116
Braunschweig, Germany }

\begin{abstract}
An all-superconductor charge qubit enabling a radio-frequency
readout of its quantum state is described. The core element of
the setup is a superconducting loop which includes the
single-Cooper-pair (Bloch) transistor. This circuit has two
functions: First, it operates as a charge qubit with magnetic
control of Josephson coupling and electrostatic control of the
charge on the transistor island. Secondly, it acts as the
transducer of the rf electrometer, which probes the qubit state
by measuring the Josephson inductance of the transistor. The
evaluation of the basic parameters of this device shows its
superiority over the rf-SET-based qubit setup.

\verb PACS  numbers: 74.50.+r, 85.25.Na, 03.67.Lx
\end{abstract}
\maketitle
\newpage

Superconducting structures with small Josephson tunnel junctions
serve as a basis for electronic devices operating on single
Cooper pairs and possessing remarkable characteristics. The paper
Ref.~\cite{QBit} has demonstrated the potential of the
single-Cooper-pair box circuit \cite{Bouchiat} as a charge qubit
and thus has attracted renewed attention to this field. The
practical realization of the Cooper pair qubit is not, however,
simple and the main problems here are the achievement of a
reliable readout of the quantum state and an elongation of the
decoherence time. For the most part, these two issues are
interrelated because a charge detector coupled to the qubit
presents the principal source of quantum decoherence.

The qubit setup consisting of a {\it Cooper pair} box and a
capacitively coupled single {\it electron} transistor (SET,
including the rf-SET \cite{Schoel}) has been extensively explored
\cite{Makhlin,DevSch,Aassime}. Although a preliminary analysis
shows that qubit states can, in principle, be measured in the
snap-shot regime \cite{Aassime}, the `mismatch' of the charge
carriers in the setup components (namely, the incoherent nature
of charges in the SET) might lead to an unaccounted enhancement of
decoherence in the system. Alternatively, the generic type of
{\it Cooper-pair} (Bloch-transistor \cite{Bl-tr-r}) electrometers
made from superconductors \cite{Bl-electr2,Cottet} seems to be
quite promising as regards matching with the Cooper pair box.
Furthermore, similar to dc-SQUIDs \cite{Danilov} and in contrast
to SETs, the resistively shunted Cooper-pair electrometer belongs
to the category of perfect (quantum-limited) linear detectors
\cite{Zorin1,Zorin3} and, therefore, can perform continuous
measurements of a quantum object \cite{KorAv}.

Recently, we have proposed an rf-driven single-Cooper-pair
electrometer \cite{Zorin2} whose energy-resolution figure
$\epsilon$ can approach the standard quantum limit of $\hbar/2$.
The transducer of this electrometer is a Bloch transistor inserted
into a superconducting loop. The magnitude of the supercurrent
circulating in the loop depends on the polarization charge
(quasicharge) on the transistor island induced via a coupling
capacitance by the charge source, e.g., the qubit.

\begin{figure}
\begin{center}
\includegraphics[width=2.8in]{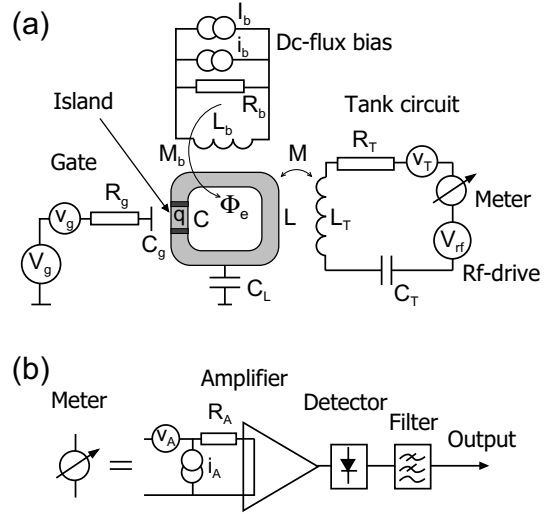}
\caption{(a) Electric diagram of the qubit-electrometer system
consisting of a macroscopic superconducting low-inductance loop
which includes two small Josephson junctions (shown by black
color) forming a mesoscopic island in between, equipped with a
capacitively coupled gate (the Bloch transistor) and a
series-resonance tank circuit inductively coupled to the loop and
driven by source $V_{\rm rf}$. The dc flux $\Phi_{\rm e}$ is
applied to the loop by a separate circuit. (b) The rf current
meter inserted into the tank circuit is a linear amplifier
followed by an amplitude (or phase) detector and a low-pass
filter.} \label{EqvSchm}
\end{center}
\end{figure}

In this paper we present a circuit in which the electrometer's
transducer takes over the function of the Cooper pair box (qubit)
as well. The device's core element is a superconducting loop
including a mesoscopic double Josephson junction with a
capacitive gate (transistor); it is shown in Fig.~1a. The
individual junctions are characterized by coupling energies
$E_{\rm J1}$ and $E_{\rm J2}$, which do not differ significantly,
\begin{equation}
k_{\rm B}T \ll\delta E_{\rm J}\equiv |E_{\rm J1}-E_{\rm J2}| \ll
E_{\rm J}\simeq E_{\rm c},
\end{equation}
where $E_{\rm J}\equiv (E_{\rm J1}+E_{\rm J2})/2$ is the average
Josephson coupling energy and $E_{\rm c} \equiv e^2/2C$ is the
total charging energy of the island (center electrode)
\cite{gap}. The island's total capacitance $C$ is much lower than
$C_{\rm L}$ (the capacitance of the `macroscopic' loop with
respect to ground). The inductance of the loop $L$ is however
rather small,
\begin{equation} \label{beta_L} \beta_{\rm L} = 2\pi L I_{\rm c}(q)/\Phi_0 \ll 1,
\end{equation}
where $\Phi_0=h/2e \approx 2.07$~fWb is the flux quantum. The
resulting critical current $I_{\rm c}$ (a function of the
island's quasicharge $q$) is always lower than its nominal value
$I_{c0}=\min\{I_{\rm c1},I_{\rm c2}\}$, where $I_{\rm c1,c2} =
(2\pi/\Phi_0)E_{\rm J1,J2}$, which is realized in the absence of
the charging effect (see, e.g., Ref.~\cite{Zorin1}).
Equation~(\ref{beta_L}) ensures a linear relation between the
overall Josephson phase, i.e. the sum of individual phases,
$\varphi = \varphi_1 + \varphi_2$ (a good quantum variable) and
the external magnetic flux $\Phi_{\rm e}$ applied to the loop.
Moreover, at small $L$ the characteristic magnetic energy of the
loop is $E_{\rm m} \equiv \Phi^2_0/2L \gg E_{\rm J}$, and, hence,
by far exceeds the energy of thermal fluctuations $k_{\rm B}T$.

The quasicharge (another good variable) \cite{LiZo}, $q = C_{\rm
g}V_{\rm g}$, is controlled by the voltage $V_{\rm g}$ applied to
a gate with coupling capacitance $C_{\rm g} \ll C$. A weakly
coupled dc-flux-bias circuit fixes the value of the frustration
parameter $f = \Phi_{\rm e}/\Phi_0$ and, hence, the dc Josephson
phase $\varphi_0=2\pi f$. In particular, the value of $f = 0.5$
makes it possible to significantly reduce the effective Josephson
coupling $\tilde{E}_{\rm J}$ of the system as a Cooper-pair box
(qubit), i.e. $\tilde{E}_{\rm J}= (E^2_{\rm J1}+E^2_{\rm
J2}+2E_{\rm J1}E_{\rm J2}\cos \varphi_0)^{1/2}= \delta E_{\rm J}$,
and thereby to localize the region, $q\approx e$, of intensive
mixing of the island's charge states $|0\rangle$ and
$|2e\rangle$, resulting in the symmetric and antisymmetric basis
states: $\frac{1}{\sqrt{2}}(|0\rangle +|2e\rangle)$ at $n=0$ and
$\frac{1}{\sqrt{2}}(|0\rangle -|2e\rangle)$ at $n=1$ respectively.
The corresponding energy spectrum (two lowest Bloch bands)
\cite{LiZo} is presented in Fig.~2.

\begin{figure}
\begin{center}
\includegraphics[width = 2.8in]{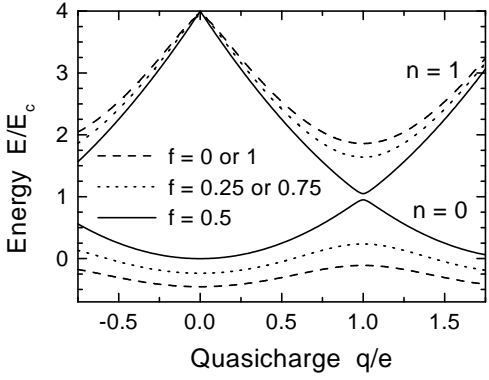}
\caption{Energy of the ground ($n=0$) and upper ($n=1$) states of
the Bloch transistor inserted into the low-inductance
superconducting loop at different values of frustration parameter
$f$. The resultant Josephson coupling of the Cooper pair box
determines the energy gap in the degeneracy point $q = e$ and
varies from $2E_{\rm J}$ (at $f=0$) to $\delta E_{\rm J}$ (at
$f=0.5$). The circuit parameters are: $E_{\rm J} = E_{\rm c}$ and
$\delta E_{\rm J} = 0.1\,E_{\rm J}$.} \label{EIc}
\end{center}
\end{figure}

Finally, as shown in the diagram of Fig.~1a, the loop is
inductively coupled to a high-$Q$ series tank circuit driven by
the sinusoidal voltage $V_{\rm rf}$ of frequency $\omega \approx
\omega_0 \equiv (L_{\rm T} C_{\rm T})^{-1/2} \ll \delta E_{\rm
J}/\hbar$. Mutual coupling is relatively weak, $k=M/(LL_{\rm
T})^{1/2} \ll 1$, while the product $k^2Q\beta_{\rm L}$ is not
small $(\sim 1)$. The change of the amplitude of current
oscillations in the tank is amplified, detected, filtered and
then serves as an output signal (see Fig.~1b).

The regime analyzed in Ref.~\cite{Zorin2} assumes a substantial
amplitude of induced oscillations of the Josephson phase,
$\varphi = \varphi_0 + a \sin\omega t$ with $a \approx 1.8$. This
value yields maximum output signals of the rf Bloch electrometer
(as well as of the single-junction SQUID operating in the similar,
non-hysteretic mode \cite{Hansma}). In this regime, phase
$\varphi$ spans the whole period of $2\pi$; so the oscillations
in the tank in fact probe the critical current of the transistor
$I_{\rm c}$, whose value depends on the polarization charge on
its island $q$. This charge can be induced, for example, by a
charge of a standalone Cooper-pair box coupled to such
electrometer via a small capacitance.

In the present case of `100\% coupling' between box and
electrometer, the amplitude of driving signal $V_{\rm rf}$ can be
reduced so that $a \ll 1$. In this regime, the impedance of the
low-$\beta_{\rm L}$ loop with the Bloch transistor is determined
by the Josephson inductance $L_{\rm J}$ whose reverse value is
equal to
\begin{equation} \label{L_J} L_{\rm J}^{-1}(q,n) =
\frac{2\pi}{\Phi_0} \frac{\partial I_{\rm s}(\varphi,
q,n)}{\partial \varphi}.
\end{equation}
Due to coupling to the loop the effective inductance of the tank
is changed,
\begin{equation} \label{L_eff} L_{\rm T} \,\rightarrow \, L_{\rm T}-
M^2 L^{-1}_{\rm J}(q,n),
\end{equation}
and this leads to the shift of the resonance frequency,
$\delta\omega_0/\omega_0 \approx k^2 L L^{-1}_{\rm J}/2$
\cite{RifIl}.

\begin{figure}
\begin{center}
\includegraphics[width = 2.8in]{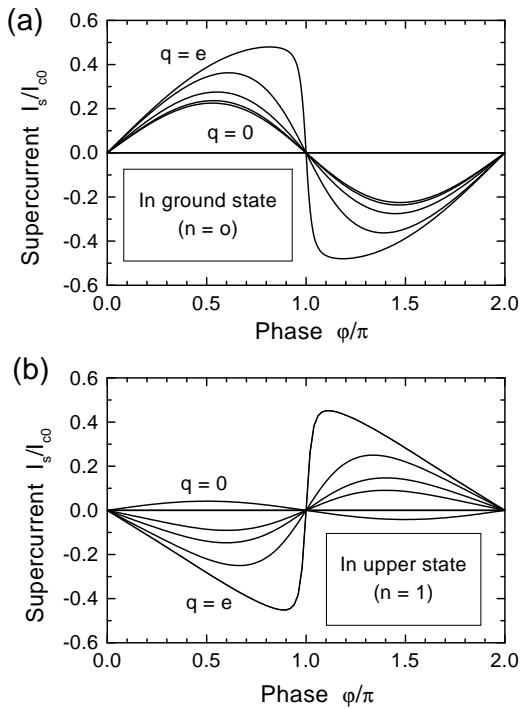}
\caption{Supercurrent-phase relation in the Bloch transistor with
$E_{\rm J} = E_{\rm c}$ and $\delta E_{\rm J} = 0.1\,E_{\rm J}$
in the ground state \cite{Zorin3} (a) and in the upper state (b)
calculated in the points of $q = 0, 0.25e, 0.5e, 0.75e$ and $e$
(in the order of increasing amplitude). Note the transformation
of the current-phase relation from the regular one to the
`$\pi$-shifted' relation, occurring at small $q$ in the upper
Bloch state (b).} \label{IsFi}
\end{center}
\end{figure}

To evaluate this shift we compute the expectation value of the
supercurrent operator $\hat{I}_{\rm s}$ using the Bloch
eigenfunctions $|q,n\rangle$ \cite{LiZo} as follows
\cite{Bl-tr-r},
\begin{eqnarray} \label{I_sup} I_{\rm s}(\varphi,q,n)
= \langle q,n| \hat{I}_{\rm s}|q,n\rangle 
= \langle q,n| I_{\rm c1} \sin \varphi_1|q,n\rangle \nonumber \\
= \langle q,n| I_{\rm c2} \sin \varphi_2|q,n\rangle.
\end{eqnarray}
The result of numerical calculation for the case of almost
symmetric transistor with $E_{\rm J} = E_{\rm c}$ is presented in
Fig.~3. As can be seen from these plots, not only the maximum of
Josephson supercurrent, i.e. the critical current value, depends
on the quasicharge $q$ and the band index $n$ \cite{Flees}, but
also its phase dependence for $n=0$ and $n=1$ is different. In
the ground state the current-phase dependence has a shape typical
of the Josephson weak links (see, e.g., the review paper
\cite{weaklinks} and references therein). In the upper state, for
all $q$ not close to 0 mod$(2e)$ and not very large values of the
ratio $E_{\rm J}/E_{\rm c}$ the dependence has a phase shift of
$\pi$, which is typical of the Josephson junctions with
ferromagnetic interlayer \cite{pi-j}.

This behavior of the supercurrent results in the strong
dependence of the Josephson inductance (see Eq.~(\ref{L_J})) on
$q$ and $n$ at $\varphi_0 \approx \pi$, viz., $L_{\rm J}^{-1}$ is
negative in the ground state and positive in the upper state,
while its absolute values $\sim (I_{\rm c0}/\Phi_0)$. Therefore,
when the band number $n$ is changing, $0\leftrightarrows 1$, the
relative shift of the resonance frequency is substantial, i.e.,
\begin{equation}\delta\omega_0/\omega_0
\sim k^2 \beta_{\rm L}\sim Q^{-1}.
\end{equation}
Near the resonance this shift leads to a dramatic change of the
amplitude of ac current in the tank, $\delta I_{\rm a} \sim
I_{\rm a} = (\Phi_0/2\pi M)\,a$, that is reliably measured at $a
\gtrsim (\Theta_{\rm A}B/E_{\rm J}\omega)^{1/2}$, assuming the
output bandwidth $B \lesssim \omega/Q \ll \omega$ and dominating
role of the noise of the amplifier whose noise temperature
$T_{\rm A} \equiv \Theta_{\rm A}/k_{\rm B}\gg T$ and the input
impedance $R_{\rm A}\gg R_{\rm T}$ \cite{Mueck}.

The great advantage of the superconducting loop including the
Bloch transistor consists in a negligibly low dissipation and,
therefore, a minor role of intrinsic sources of qubit decoherence
\cite{Q0}. The processes associated with dissipation are the
quasiparticle and the pair-quasiparticle interference
\cite{Langenberg} components of the tunneling. In our case,
however, both the frequency and the amplitude of the voltage
across the transistor $V_{\rm tr}$ are small, namely $2eV_{\rm
tr}= a \hbar \omega \ll \hbar\omega \lesssim k_{\rm B}T \ll
E_{\rm c} < \Delta$. This relation ensures huge suppression (by a
factor $\eta = \exp (-\Delta/k_{\rm B}T) \lll 1$) of all
dissipative processes. Moreover, since the parity effect blocks
sequential quasiparticle tunneling, dissipation can only occur
due to the co-tunneling effect. The effective leakage resistance
of the transistor (with normal resistances of the junctions $R_1$
and $R_2$) therefore is of the order of $(R_1 R_2/R_{\rm
q})(\Delta/a \eta \hbar \omega)^2 \ggg R_{1,2}$, where $R_{\rm
q}=h/4e^2$ is the resistance quantum, and can, therefore, be
neglected.

The main source of decoherence therefore is the external
electromagnetic circuit: it causes fluctuations of the island's
electric potential $\tilde{u}$. The fluctuation sources, labeled
as $v_{\rm g}$, $i_{\rm b}$, $v_{\rm T}$ and $v_{\rm A}$ in
Fig.~1, are associated with dissipative components of the
circuit, $R_{\rm g}$, $R_{\rm b}$, $R_{\rm T}$ (being presumably
at the equilibrium temperature $T$) and $R_{\rm A}$
(characterized by $T_{\rm A}$). The rates of dephasing caused by
the gate and dc-flux bias, i.e. by the sources $v_{\rm g}$ and
$i_{\rm b}$ respectively, were earlier evaluated by Makhlin, et
al. \cite {Makhlin}. They showed that reduction of the coupling
strengths, i.e., $C_{\rm g}$ and $M_{\rm b}$, can make these
rates small enough to allow many single-bit manipulations to be
performed within the dephasing time. If $V_{\rm rf}=0$, similar
reasoning can be applied to the resonance tank circuit containing
the sources $v_{\rm T}$ and $v_{\rm A}$.

When the rf-drive is on, it leads to an enhancement of the
fluctuations $\tilde{u}$ at low frequencies $(\ll \omega)$ due to
parametric down-conversion of noise at frequencies around
$\omega$. The spectral density of these fluctuations
\cite{Zorin2},
\begin{equation}S_{\rm u}(0)\approx a^2 E_{\rm J} \Theta_{\rm
A}/e^2 \omega,
\end{equation}
is equivalent to that of the resistance $R_{\rm eff}=a^2(E_{\rm
J}/\hbar\omega)R_{\rm q}$ at temperature $T_{\rm A}$. At $T_{\rm
A} \approx \hbar\omega/k_{\rm B}$, the resistance $R_{\rm eff}
\sim R_{\rm q}/Q$, which allows \cite{Makhlin} up to $N = R_{\rm
q}/R_{\rm eff} \sim Q \:({\rm say, }\sim 10^3)$ single-qubit
manipulations to be performed in the degeneracy point, $q=e$,
during the entanglement time $\tau_{\rm ent} \sim Q\hbar/\delta
E_{\rm J}$. In contrast to the rf-SET setup \cite{Aassime}, there
is, in principle, no need for switching the electrometer off in
the degeneracy point $q=e$.

During the measurement (away from the point $q=e$) the energy gap
between the charge states is large, $\Delta E \equiv \hbar\Omega
\gtrsim E_{\rm c} > \delta E_{\rm J}$. Due to the high impedance
of the tank at the transition frequency $\Omega \gg\omega_0$, the
spectral density $S_{\rm u}(\Omega)$ is remarkably low, viz.
$\approx \hbar\omega/\pi Q\Omega L_{\rm T}$, which yields a long
mixing time \cite{DevSch} $\tau_{\rm mix}=(\hbar/e)^2 (\Delta
E/E_{\rm J})^2 S_{\rm u}^{-2}(\Omega)$. The latter value allows a
measurement with a high signal-to-noise ratio,
\begin{equation}S/N = (B
\tau_{\rm mix})^{1/2} \sim (\hbar\Omega/k_{\rm B}T_{\rm
A})^{1/2}Q \gg 1, \end{equation} to be performed even if a cooled
semiconductor amplifier with the fairly good noise temperature of
$T_{\rm A}\sim 10$~K is exploited (cf. $S/N \approx 4$ achievable
using an Al rf-SET with extremely low noise figure
\cite{Aassime}).

I am grateful to M.~G\"otz, M.~M\"uck, and J.~Niemeyer for
valuable discussions.


\begin{thebibliography}{00}

\bibitem{QBit} Y.~Nakamura, Y.~A.~Pashkin, and J.~S.~Tsai,
Nature 398 (1999) 768.

\bibitem{Bouchiat}
V.~Bouchiat, D.~Vion, P.~Joyez, D.~Esteve, and M.~H.~Devoret,
Phys. Scr. T76 (1998) 165.

\bibitem{Schoel} R.~J.~Schoelkopf, P.~Wahlgren, A.~A.~Kozhevnikov,
P.~Delsing, and D.~E.~Prober, Science 280 (1998) 1238.

\bibitem{Makhlin} Yu.~Makhlin, G.~Sch\"on, and A.~Shnirman,
Rev. Mod. Phys. 73 (2001) 357.

\bibitem{DevSch} M.~H.~Devoret and R.~J.~Schoelkopf, Nature
406 (2000) 1039.

\bibitem{Aassime} A.~Aassime, G.~Johansson, G.~Wendin, R.~J.~Schoelkopf,
and P.~Delsing, Phys. Rev. Lett. 86 (2001) 3376.

\bibitem{Bl-tr-r} K.~K.~Likharev, Moscow State University Preprint
No.29 (1986); D.~V.~Averin and K.~K.~Likharev, in: Mesoscopic
Phenomena in Solids, edited by B.~L.~Altshuler, P.~A.~Lee, and
R.~A.~Webb (Elsevier, Amsterdam, 1991) p.~213.

\bibitem{Bl-electr2} S.~V.~Lotkhov,  H.~Zangerle, A.~B.~Zorin, T.~Weimann,
H.~Scherer, and J.~Niemeyer,  IEEE Trans. Appl. Supercond. 9
(1999) 3664; A.~B.~Zorin, S.~V.~Lotkhov, Yu.~A.~Pashkin,
H.~Zangerle, V.~A.~Krupenin, T.~Weimann, H.~Scherer, and
J.~Niemeyer, J. Supercond. 12 (1999) 747.

\bibitem{Cottet} A.~Cottet, A.~Steinbach, P.~Joyez, D.~Vion,
H.~Pothier, D.~Esteve, and M.~E.~Huber, in: Macroscopic Quantum
Tunneling and Quantum Computing, edited by D.~V.~Averin,
B.~Ruggiero, and P.~Silvestrini (Kluwer, New York, 2001), p. 111.

\bibitem{Danilov} V.~V.~Danilov, K.~K.~Likharev, and A.~B.~Zorin,
IEEE Trans. Magn. 19 (1983) 572.

\bibitem{Zorin1} A.~B.~Zorin, Phys. Rev. Lett. 76 (1996)
4408.

\bibitem{Zorin3} A.~B.~Zorin, IEEE Trans. Instrum. and Meas.
46 (1997) 299.

\bibitem{KorAv} A.~N.~Korotkov, Phys. Rev. B 63 (2001) 085312;
D.~V.~Averin, cond-mat/0004364.

\bibitem{Zorin2} A.~B.~Zorin, Phys. Rev. Lett. 86 (2001) 3388.

\bibitem{gap} We assume that the superconductor energy gap
$\Delta$ exceeds $E_{\rm c}$, so the quasiparticle tunneling is
suppressed due to the parity effect (see M.~T.~Tuominen,
J.~M.~Hergenrother, T.~S.~Tighe, and M.~Tinkham, Phys. Rev. Lett.
69 (1992) 1997). In the case of Al electrodes $\Delta_{\rm Al}
\approx 200~\mu$eV, so the energies $E_{\rm c}$ and $E_{\rm J}$
can be of the order of 100~$\mu$eV and temperature $T < 100$~mK.

\bibitem{LiZo} K.~K.~Likharev and A.~B.~Zorin, J. Low Temp. Phys.
59 (1985) 347; D.~V.~Averin, A.~B.~Zorin, and K.~K.~Likharev,
Sov. Phys. JETP 88 (1985) 697.

\bibitem{Hansma} P.~K.~Hansma, J. Appl. Phys. 44 (1973) 4191.

\bibitem{RifIl} A similar measurement scheme in a traditional
single-junction rf-SQUID was proposed by R.~Rifkin and
B.~S.~Deaver, Jr., Phys. Rev. B 13 (1976) 3894. Lately, the
efficiency of this method (in somewhat modified version) was
demonstrated in measurement of the phase dependence of a
tens-of-nanoampere supercurrent in small Nb double junctions
($E_{\rm J} \approx 300~\mu$eV) at temperature $T=4.2$~K by
E.~Il'ichev, V.~Zakosarenko, L.~Fritzsch, R.~Stolz, H.~E.~Hoenig,
H.-G.~Meyer, M.~G\"otz, A.~B.~Zorin, V.~V.~Khanin,
A.~B.~Pavolotsky, and J.~Niemeyer, Rev. Sci. Instrum. 72 (2001)
1882.

\bibitem{Flees} D.~J.~Flees, S.~Han, and J.~E.~Lukens, Phys. Rev. Lett.
78 (1997) 4817.

\bibitem{weaklinks} K.~K.~Likharev, Rev. Mod. Phys. 51 (1979) 101.

\bibitem{pi-j} V.~V.~Ryazanov, V.~A.~Oboznov, A.~Yu.~Rusanov,
A.~V.~Veretennikov, A.~A.~Golubov, and J.~Aarts, Phys. Rev. Lett.
86 (2001) 2427.


\bibitem{Mueck} In the state-of-the-art dc-SQUID amplifiers for the
frequency band 100\,MHz$-$1\,GHz the noise figure $\Theta_{\rm
A}$ is of the order of $\hbar\omega$, see M.~M\"uck, J.~B.~Kycia,
and J.~Clarke, Appl. Phys. Lett. 78 (2001) 967 and references
therein.

\bibitem{Q0} We do not address here the effect of
background charge fluctuations on the qubit island, whose strength
strongly depends on the sample design and operating regime (see
V.~A.~Krupenin, D.~E.~Presnov, M.~N.~Savvateev, H.~Scherer,
A.~B.~Zorin, and J.~Niemeyer, J. Appl. Phys. 48 (1998) 3212) and
may present a problem in experiment (see, e.g., the recent paper
on probing the box-transistor states by single electron
tunneling: Y.~Nakamura, Yu.~A.~Pashkin, T.~Yamamoto, and
J.~S.~Tsai, Preprint (2001)).

\bibitem{Langenberg} D.~N.~Langenberg, Rev. Phys. Appl. 9 (1974) 35.


\end{thebibliography}
\end{document}